\documentclass[prd,a4paper,showpacs,showkeys,preprint]{revtex4}
\usepackage{graphicx}
\usepackage{dcolumn}

\begin{document}

\title{A unified meson-baryon potential}
\author{Fabian \surname{Brau}}
\thanks{FNRS Postdoctoral Researcher}
\email[E-mail: ]{fabian.brau@umh.ac.be}
\author{Claude \surname{Semay}}
\thanks{FNRS Research Associate}
\email[E-mail: ]{claude.semay@umh.ac.be}
\affiliation{Universit\'{e} de Mons-Hainaut, Place du Parc, 20,
B-7000 Mons, Belgium}
\author{Bernard \surname{Silvestre-Brac}}
\email[E-mail: ]{silvestre@isn.in2p3.fr}
\affiliation{Institut des Sciences Nucl\'{e}aires,
53, Av. des Martyrs, F-38026 Grenoble-Cedex, France}
\date{\today}

\begin{abstract}
We study the spectra of mesons and baryons, composed of light quarks, in
the framework of a semirelativistic potential model including instanton
induced forces. We show how a simple modification of the instanton
interaction in the baryon sector allows a good description of the meson
and the baryon spectra using an interaction characterized by a unique
set of parameters.
\end{abstract}

\pacs{12.39.Pn, 14.20.-c}
\keywords{Relativistic quark model; Potential models;
Mesons; Baryons}

\maketitle

\section{Introduction}
\label{sec:intro}

The description of the meson and baryon spectra in the framework of
nonrelativistic or semirelativistic potential models appears to be a
rather successful approach. Many works have been devoted to the study of
these observables but generally not in a consistent way: the
meson properties (see for example
Refs.~\cite{junc92,luch91,fulc94,sema97,blan99}) or the baryon
properties (see for
example Refs.~\cite{silv85,stan90,desp92,gloz98}) are investigated in
disconnected
approaches. The spectra are generally well reproduced separately; only
the understanding of few states remains difficult (radial excitations of
kaons or $\Lambda(1405)$, for example). Nevertheless a unified
description of meson and baryon spectra seems to be more problematic.
For example, in Ref.~\cite{sema97}, the meson spectra are nicely
reproduced, but the baryon spectra with the same potential are not so
satisfactory. Conversely, the model of Ref.~\cite{gloz98} is rather good
for baryon spectra, but appears catastrophic if applied, as such, to
describe the
meson properties. There exist only few complete studies dealing
consistently with both meson and baryon spectra. Even if some
encouraging results have already been obtained, none is really
satisfying.

In a pioneer work, Bhaduri {\it et al.} have proposed a nonrelativistic
model relying on a Cornell potential and a spin-spin interaction
\cite{bhad81}. The authors proposed a consistent scheme and used the
same set of parameters for mesons and baryons. This model was refined
later on by Silvestre-Brac and Semay \cite{silv93} with interesting
successes. However, the main problem arising with these
models is the bad description of the radial excitations of mesons and
baryons; this is due partly to the use of nonrelativistic kinematics
\cite{brau00}. Most of these excitations are calculated 200-300 MeV
above their experimental value. Moreover, pseudoscalar mesons cannot
be described in a satisfactory way since the interaction does not allow
flavor mixing and thus does not allow a correct description of the
mesons $\eta$ and $\eta'$.

Isgur {\it et al.} have proposed a semirelativistic model
relying on a Y junction for a confinement supplemented by spin-spin,
spin-orbit, and tensor interactions \cite{isgu86} (denoted CGI model in
this paper). A purely phenomenological annihilation term was added to
describe the flavor mixing. The authors also proposed to take into
account some relativistic effects by replacing the masses of the quarks
appearing in the interaction by expressions which depend on quark
momenta; this procedure introduces new phenomenological parameters. This
leads to a complicated model which, up to our knowledge, has never been
used for systems containing more than three quarks. Despite the use of
such a complex model, the authors need to choose different values for
the slope of the confinement for mesons and for baryons to obtain good
theoretical results (some other parameters have also been slightly
modified). The meson spectra obtained with this model are good and
similar to spectra obtained with more simple models (see for example
Ref.~\cite{brau98}). The baryon spectra are also in good agreement with
experiment (but less good than the meson spectra) and here also similar
to spectra
obtained with more simple models (see for example Ref.~\cite{sema01}).
Note that the baryon spectra are clearly less good than the ones
obtained with Goldstone Boson-Exchange models \cite{gloz98}, but a
unified description of mesons and baryons
seems to be difficult within this formalism \cite{isgu00}.

Another attempt to get a consistent description of meson and baryon
properties with a simple model was performed by Blask {\it et al.} \cite
{blas90} (denoted BBHMP model in this paper). The authors proposed a
nonrelativistic model relying on a confinement
supplemented by an instanton induced interaction \cite{shur89}. The
meson and baryon
ground states are well described but the use of a nonrelativistic
kinematics leads, here also, to a bad description of radial excitations
of mesons and baryons.

These previous works devoted to the description of meson and baryon
properties with a unique interaction show clearly that this task is
complex, and no satisfying result has already been obtained. Very
recently, L\"{o}ring {\it et al.} have studied meson and baryon spectra
within a relativistic framework based on a covariant Bethe-Salpeter
equation \cite{lori01}. Even with this more sophisticated model an
unified description is not possible.

In a previous work \cite{sema01}, we tried such a description using a
semirelativistic model relying on a Cornell potential supplemented by an
instanton induced interaction. No satisfying result was obtained, but
this work \cite{sema01} and a previous one \cite{brau98} have clearly
shown that a separate description of meson and baryon properties was
possible within this simple semirelativistic potential. So, a natural
question arises: ``which simple modification (if any) of our model could
lead to a consistent description of both meson and baryon spectra?".

In this work, we show that, at least one such a modification
exists. The one we present consists in adding a simple constant term in
the instanton induced interaction in the baryon sector.

\section{Model}
\label{sec:model}

The model used in this work is similar to those introduced in
Refs.~\cite{brau98,sema01}, except from the additional mentioned term in
the instanton induced interaction (see below). Here, we just recall the
main lines of our model.

The Hamiltonian is written
\begin{equation}
\label{hamil}
H= \sum_{i=1}^{N} \sqrt{{\vec p \,}_i^2 + m_i^2} + \sum_{i<j=1}^{N}
V_{ij} \quad (N = 2 \ \text{or}\ 3),
\end{equation}
with $\vec p_i$ the momentum of quark $i$
($\sum_{i=1}^{N} \vec p_i = \vec 0$), $m_i$ its constituent mass, and
$V_{ij}$ the interaction between quarks (or antiquarks) $i$ and $j$. The
interaction contains the Cornell potential and the instanton induced
interaction. The Cornell potential, which depends only on the distance
$r$ between two quarks, is given by
\begin{equation}
\label{cornell}
V_{\text{C}}(r) = -\frac{3}{4} \frac{\lambda_i\cdot\lambda_j}{4}\left[ -
\frac{\kappa}{r}+a\, r+C_{\text{M}}
+C_{\text{B}}\delta_{N3} \right],
\end{equation}
The confining part of this potential represents a good approximation of
the string junction in a meson and of a Y-shape string configuration in
a baryon. As usual, we need two different
constant interactions to obtain correct absolute values of the meson and
baryon energy levels. $C_{\text{M}}$ is the constant for the meson
spectra, while $C_{\text{B}}$ is the constant that we need to add to
$C_{\text{M}}$ to obtain the correct absolute value of the energy levels
of the baryon spectra. The presence of the $C_{\text{B}}$ term could
simulate the effect of three-body forces. Obviously, the values of these
constants do not influence the structure of the wave function and thus
play a minor role: only the relative positions of the energy levels have
a physical meaning.

The instanton induced interaction provides a suitable formalism to
reproduce well the spectrum of the pseudoscalar mesons (and to explain
the masses  of $\eta$ and $\eta'$ mesons). In the nonrelativistic limit,
this interaction between one quark and one antiquark in a meson
\cite{blas90,munz94} is vanishing for $L\neq 0$ or $S \neq 0$ states.
For $L=S=0$, its form depends on the isospin of the $q\bar q$ pair
\begin{itemize}
\item For $I=1$:
\begin{equation}
\label{ins2}
V_{\text{I}}(r)=-8\, g\, \delta(\vec{r}\,);
\end{equation}
\item For $I=1/2$:
\begin{equation}
\label{ins3}
V_{\text{I}}(r)=-8\, g'\, \delta(\vec{r}\,);
\end{equation}
\item $I=0$:
\begin{equation}
\label{ins4}
V_{\text{I}}(r)=8
\left(
\begin{array}{cc}
g & \sqrt{2}g' \\
\sqrt{2}g' & 0
\end{array}
\right)\, \delta(\vec{r}\,),
\end{equation}
in the flavor space $(1/\sqrt{2}(|u\bar{u} \rangle+|d\bar{d}
\rangle),|s\bar{s} \rangle)$.
\end{itemize}
The parameters $g$ and $g'$ are two
dimensioned coupling constants. Between two quarks in a baryon, this
interaction is written \cite{blas90,munz94}
\begin{equation}
\label{vinst}
V_{\text{I}}(r)=-4 \left(g P^{[nn]} + g' P^{[ns]} \right) P^{S=0}
\delta(\vec r\,),
\end{equation}
where $P^{S=0}$ is the projector on spin 0, and $P^{[qq']}$ is the
projector on antisymmetrical flavor state $qq'$ ($n$ for $u$ or $d$ is a
non-strange quark, and $s$ is the strange quark). The operator
$P^{[nn]}$ is simply a projector on isosinglet states. A procedure to
compute the matrix elements of the projector $P^{[ns]}$ is described in
Ref.~\cite{sema01}.

The instanton induced forces also give a contribution $\Delta m_q$ to
the current quark mass $m^0_q$. As this interaction is not necessarily
the only source for the constituent mass, a phenomenological term
$\delta_q$ is also added to the current mass \cite{brau98}. Finally, the
constituent masses in our model are given by
\begin{eqnarray}
m_n &=& m_n^0+\Delta m_n+\delta_n, \\
m_s &=& m_s^0+\Delta m_s+\delta_s.
\end{eqnarray}
In the instanton theory, the quantities $g$, $g'$, $\Delta m_n$,
$\Delta m_s$ are given by integrals over the instanton size $\rho$ up to
a cutoff value $\rho_c$ (see for instance
Ref.~\cite[formulas~(5)-(9)]{munz94}). These integrals can be rewritten
in a more interesting form for numerical calculations by defining a
dimensionless instanton size $x=\rho\Lambda$, where $\Lambda$ is the QCD
scale parameter \cite{brau98}.

The quark masses used in our model are the constituent masses and not
the current ones. It is then natural to suppose that a quark is not a
pure point-like particle, but an effective degree of freedom which is
dressed by the gluon and quark-antiquark pair clouds. The form that we
retain for the color charge density of a quark is a Gaussian function
\begin{equation}
\label{rhoi}
\rho(\vec r\,) = \frac{1}{(\gamma\sqrt{\pi})^{3/2}} \exp(- r^2 /
\gamma^2).
\end{equation}
It is generally assumed that the quark size $\gamma$ depends on the
flavor. So, we consider two size parameters $\gamma_n$ and $\gamma_s$
for $n$ and $s$ quarks respectively. It is assumed that the dressed
expression $\widetilde O_{ij}(\vec r\,)$ of a bare operator $O_{ij}(\vec
r\,)$, which depends only on the relative distance $\vec r = \vec r_i -
\vec r_j$ between the quarks $i$ and $j$, is given by
\begin{equation}
\label{od}
\widetilde O_{ij}(\vec r\,) = \int d\vec r\,'\,O_{ij}(\vec r\,')
\rho_{ij}
(\vec r - \vec r\,'),
\end{equation}
where $\rho_{ij}$ is also a Gaussian function of type~(\ref{rhoi})
with the size parameter $\gamma_{ij}$ given by
\begin{equation}
\label{gij}
\gamma_{ij}= \sqrt{\gamma_i^2 + \gamma_j^2}.
\end{equation}
This formula is chosen because the convolution of two Gaussian
functions, with size parameters $\gamma_i$ and $\gamma_j$ respectively,
is also a Gaussian function with a size parameter given by
Eq.~(\ref{gij}) (for more details, see Ref.~\cite{brau98}).

After convolution with the quark density, a Cornell dressed
potential has the following form
\begin{equation}
\label{vfund}
- \frac{\kappa}{r} + a\,r + C \rightarrow
-\kappa \frac{\text{erf}(r/\gamma_{ij})}{r}
+ a\,r \left[\frac{\gamma_{ij}\,\exp(-r^2/\gamma_{ij}^2)}{\sqrt{\pi}
\,r} + \left( 1+ \frac{\gamma_{ij}^2}{2r^2} \right)
\text{erf}(r/\gamma_{ij}) \right] + C,
\end{equation}
while the Dirac distribution in $V_{\text{I}}(r)$ is transformed into a
Gaussian function
\begin{equation}
\label{insgauss}
\delta(\vec r\,) \rightarrow \frac{1}{(\gamma_{ij}\sqrt{\pi})^3}
\exp(-r^2/\gamma_{ij}^2).
\end{equation}
Despite this convolution, we consider, for simplicity, that the
instanton induced forces act always only on $L=0$ states.

We have shown in Ref.~\cite{sema01} that this model is not able to
describe correctly meson and baryon spectra in a consistent way. We
needed to use two different sets of parameters to get a correct
description of hadron masses. So, we have performed of series of
miminizations, starting with new ranges of parameters and studying only
some classes of hadrons. We have then remarked that it was
systematically possible to reproduce the masses of all the  mesons, as
well as the masses of the baryons for which the instanton induced
interaction does not act. Consequently, we tried to modify in different
ways $V_{\text{I}}(r)$ of Eq.~(\ref{vinst}) in the baryon sector. We
propose below the simplest form that we found, which gives good results:
\begin{equation}
\label{vinstnew}
V_{\text{I}}(r)=-4 \left(g P^{[nn]} + g' P^{[ns]} \right) P^{S=0}
\tilde{\delta}(\vec r\,)+C_{\text{I}} \left(P^{[nn]} + P^{[ns]} \right)
P^{S=0} P^{L=0},
\end{equation}
In this formula, $C_{\text{I}}$ is a new constant. Due to the presence
of the projectors, this additional term will not contribute on an equal
footing for all baryon states. The status of this supplementary term is
up to now purely phenomenological. Let us note that a three-body
instanton induced
interaction exists but its contribution is vanishing in baryon
\cite{blas90,munz94}. So, the new additional term we propose cannot be
interpreted as a simulation of such a three-body interaction.

Even if we cannot provide any physical explanation for its presence in
the interaction, we believe that such an improvement of both meson and
baryon spectra (see Sec.~\ref{ssec:spectra}) is not only a question of
chance; a physical process could exist to explain the existence of this
supplementary interaction. Investigations in this direction are in
progress.

\section{Meson and baryon spectra}
\label{ssec:spectra}

In Tables~\ref{tab1} and \ref{tab2}, we give the set of meson and baryon
resonances used to fit the parameters of the model (the numerical
techniques and the fitting procedure are explained in
Refs.~\cite{brau98,sema01}). This sample is composed of 28 states taken
from the most reliable ones (18 c.o.g. of meson multiplets and 10
baryons) \cite{pdg}.

In Table~\ref{tab3}, we present the optimal values found for the
parameters of our model. We see that all the parameters of the instanton
induced interaction as well as the slope of the confinement have values
in agreement with the expected values. The constant $C_{\text{B}}$ is
very small with respect to $C_{\text{M}}$. The origin of these constants
is not clear; several attempts to
explain the presence of these quantities can be found in the literature
\cite{luch91,grom81,carl91,sema95}. Nevertheless, as mentioned in the
introduction,
the physical meaning of these constants is not crucial since their
values do not influence the wave functions of the systems, hence the
values of other observables. The values of other parameters ($\kappa$,
quark sizes, constituent masses) are less constrained but there are
close to values found in the literature. As we have already noticed in
our previous works~\cite{brau98,sema01}, the instanton induced
interaction cannot explain alone the renormalization of the current
quark masses. Its contribution is 45 MeV for the quark $n$ and 26 MeV
for the quark $s$, values that are relatively small as compared to the
constituent masses of these quarks.

In our previous work concerning only mesons \cite{brau98}, we have found
several samples of parameters giving similar spectra (see for instance
models~I to V in this reference). The same situation occurs for the
baryons \cite{sema01}. On the contrary, when mesons and baryons are
considered together, all the minimizations that we performed
produced only very similar sets of
parameters. It seems that masses of both mesons and baryons put more
severe constraints on the parameters of model.

\begin{center}
\begin{figure}
\includegraphics*[height=8cm]{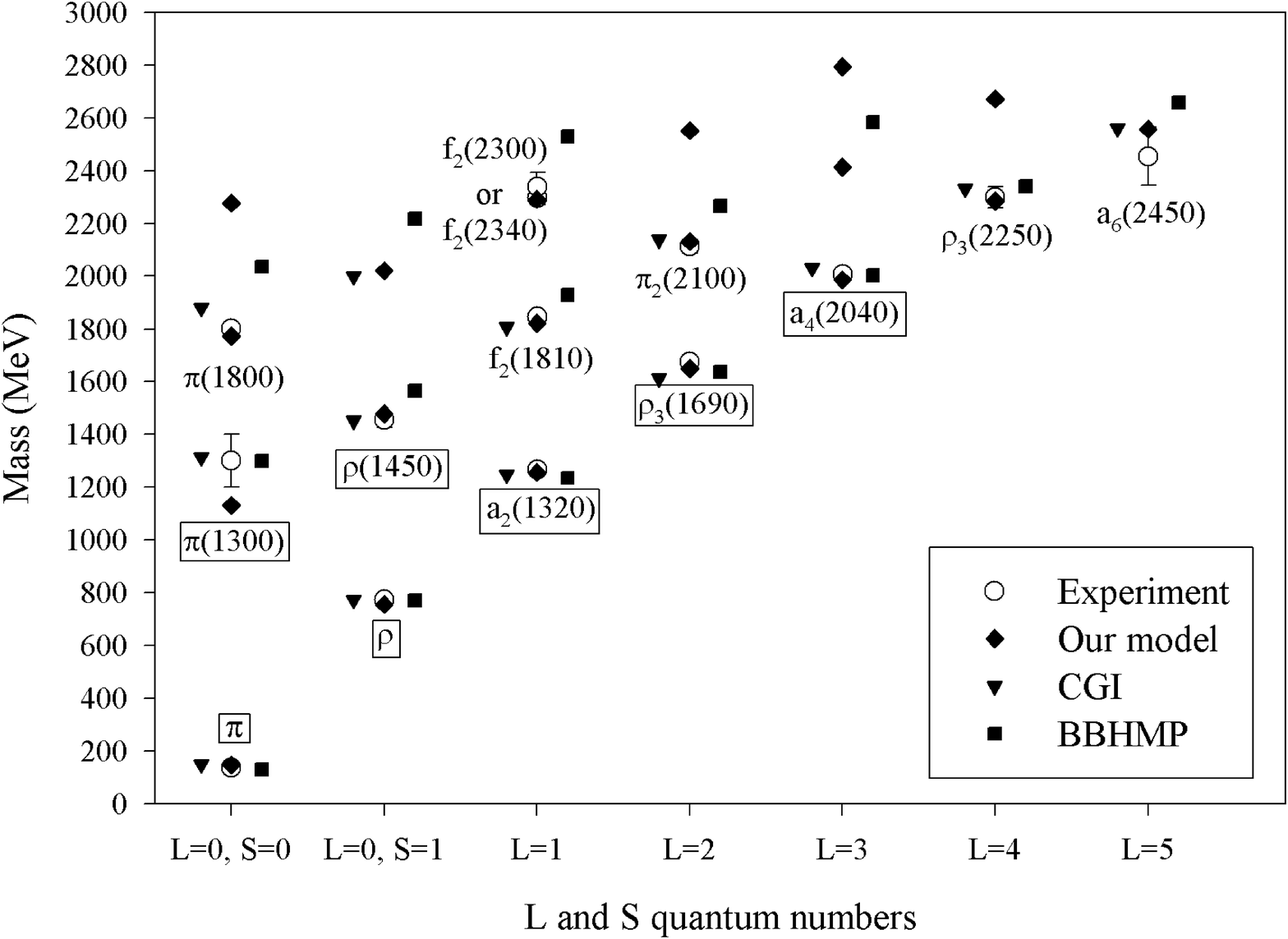}
\caption{Masses of $n\bar{n}$ mesons as a function of total orbital
angular momentum and total spin. Framed names indicate centers of
gravity of multiplets used in the fit of parameters.}
\label{fig1}
\end{figure}
\end{center}

\begin{figure}
\includegraphics*[height=8cm]{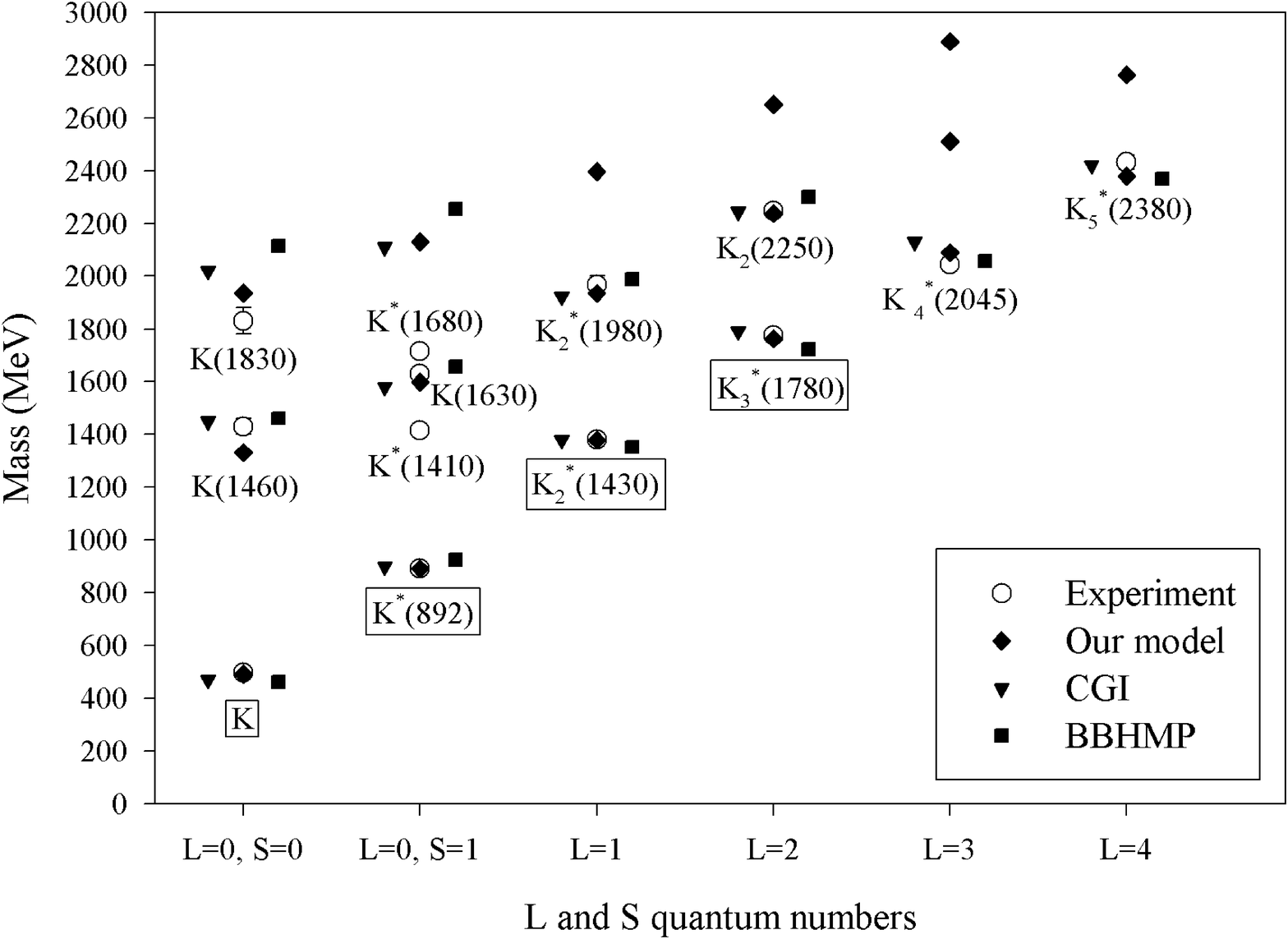}
\caption{Same as Fig.~\ref{fig1} but for $n\bar{s}$ mesons.}
\label{fig2}
\end{figure}

\begin{figure}
\includegraphics*[height=8cm]{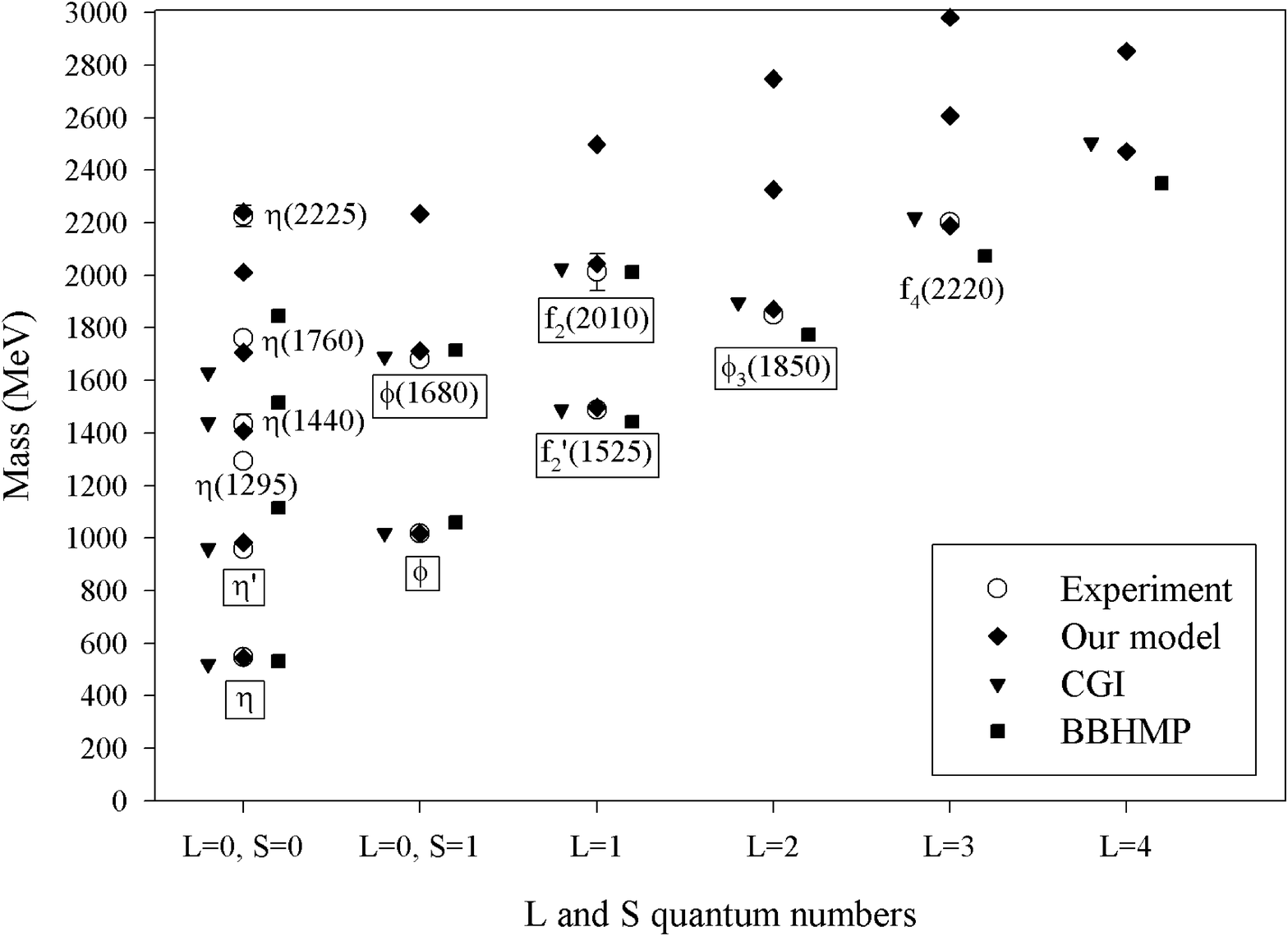}
\caption{Same as Fig.~\ref{fig1} but for $s\bar{s}$ and mixed flavor
mesons.}
\label{fig3}
\end{figure}

The Figs.~\ref{fig1}-\ref{fig3} show a comparison between the meson
spectra obtained with the model proposed in Sec.~\ref{sec:model},
the data and the results obtained with the CGI and the BBHMP models (we
do not present any comparison with the work of Bhaduri {\it et al.}
because the spectra obtained with this model have roughly the same
characteristics than the results of the BBHMP model). It is worth noting
that the
spectra obtained with our model are as good as the ones obtained in our
previous work \cite{brau98}, where the model was designed to describe
only the light mesons. Our energy levels present the same
characteristics than those found with the CGI model, but they are
clearly better than those obtained with the nonrelativistic BBHMP model
compared to experimental data. As mentioned in the introduction, the
main difficulties that appear with nonrelativistic models are the
correct description of radial excitations. Note that the pseudoscalar
mesons $\pi$-$K$-$\eta$-$\eta'$ are very well described with a unique
interaction (namely the instanton induced interaction) while in the CGI
model a purely phenomenological annihilation interaction was introduced
to describe only the $\eta$ and $\eta'$ mesons. There still exist
problems concerning radial excitations. Contrary to other works, our
model reproduces nicely the $\pi(1800)$ resonance but gives a too low
value for the $\pi(1300)$ state, which nevertheless has a very large
width. Moreover, the kaon radial excitations ($K$ and $K^*$) also
present some deficiencies. Lastly, the $\eta(1295)$ meson seems to not
fit in this scheme. One must seriously wonder whether the explanation
can be searched from a description in terms of tetraquarks or other
exotic possibilities.

\begin{figure}
\includegraphics*[height=8cm]{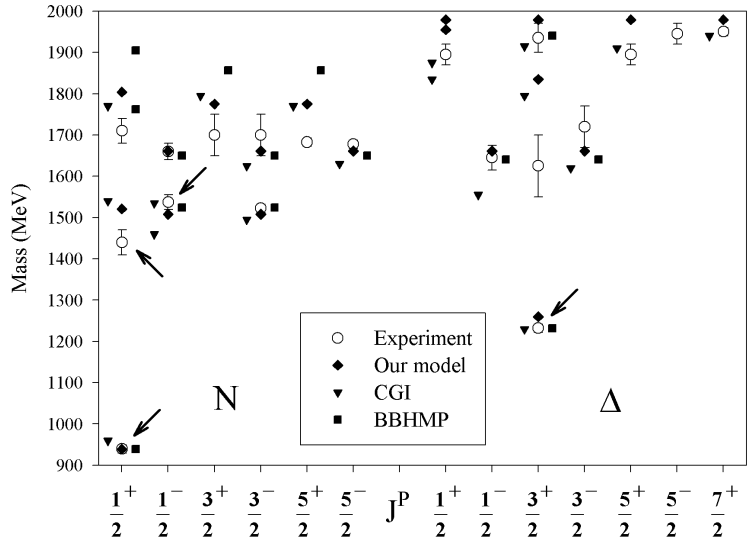}
\caption{Masses of the $N$ and $\Delta$ baryons (status
$\star$$\star$$\star$$\star$ and $\star$$\star$$\star$)
as a function of total angular momentum and parity $J^P$. The arrows
indicate baryons used in the fit of parameters.}
\label{fig4}
\end{figure}

\begin{figure}
\includegraphics*[height=8cm]{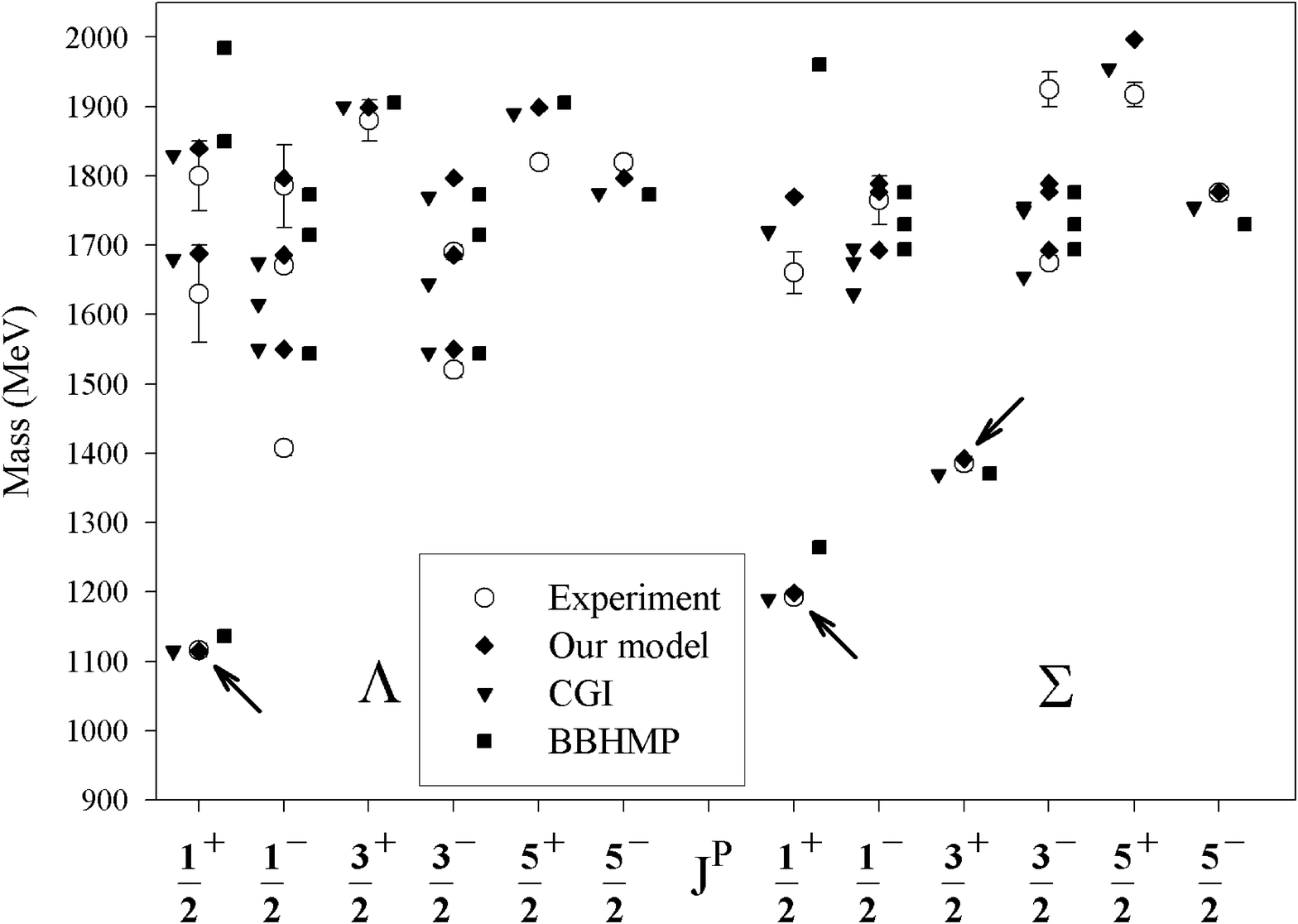}
\caption{Same as Fig.~\ref{fig4} but for the $\Lambda$ and $\Sigma$
baryons.}
\label{fig5}
\end{figure}

\begin{figure}
\includegraphics*[height=8cm]{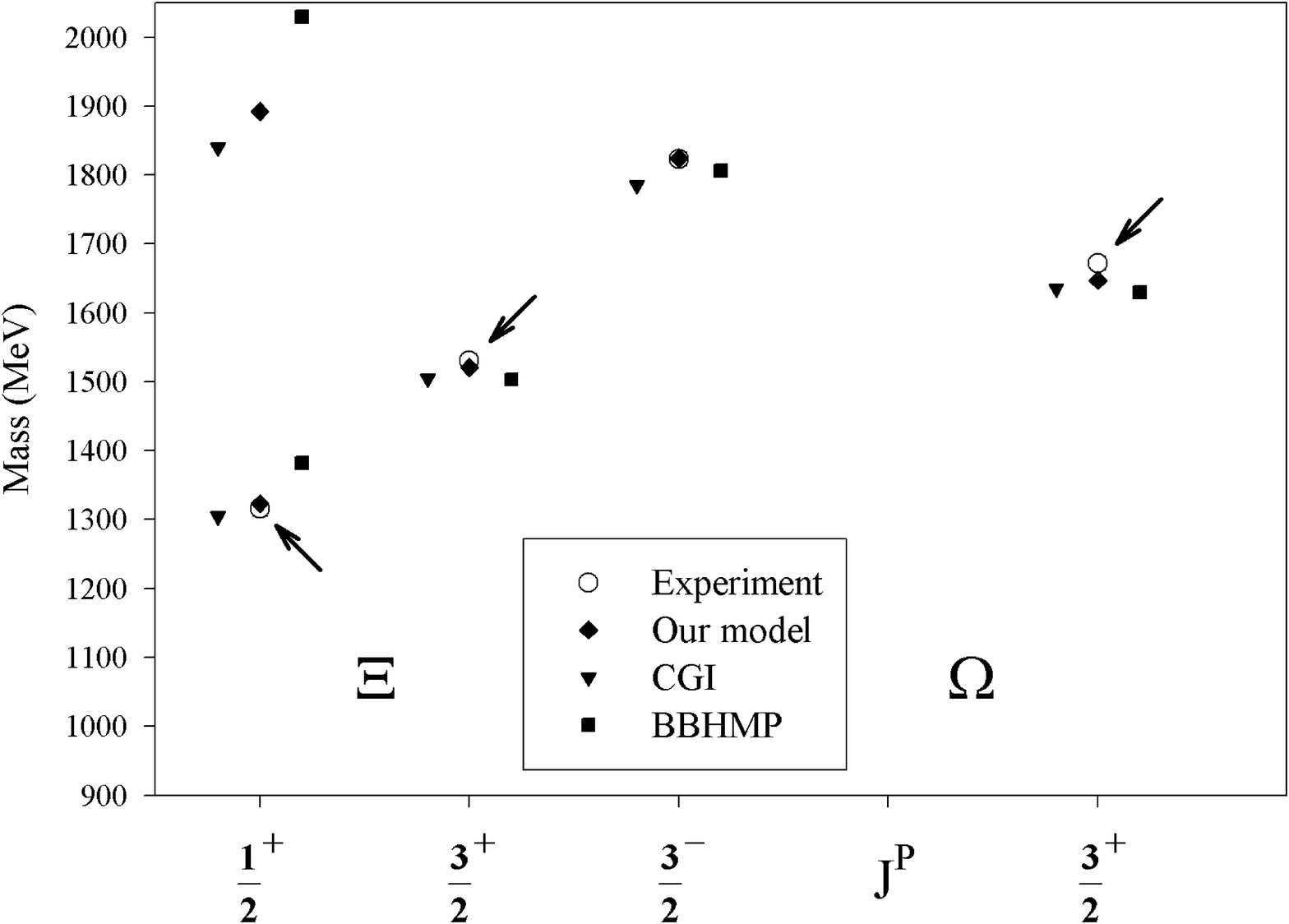}
\caption{Same as Fig.~\ref{fig4} but for the $\Xi$ and $\Omega$
baryons.}
\label{fig6}
\end{figure}

The Figs.~\ref{fig4}-\ref{fig6} present the same analysis but concerning
baryon spectra. Here also, the spectra obtained with our model are as
good as the ones obtained in our previous work \cite{sema01} where the
parameters of the model were fitted to describe only the light baryons.
Globally, the features of our spectra as compared to those obtained with
the CGI and the BBHMP models can be summarized as follow. We have an
improvement of the positive-parity states as compared to the
corresponding ones obtained with the BBHMP model, while the
negative-parity states are close in both models. Conversely, we have an
improvement of the negative-parity states as compared to the those
obtained with the CGI model while the positive-parity states are close
in both models. Although a semirelativistic kinetic energy operator
substantially improves the various Roper resonances, there still remains
a noticeable disagreement with experimental data. Moreover, the
$\Lambda(1405)$ seems to resist to any description in the framework of
potential models. It is worth mentioning that the contribution of
$q^4\bar q$ configurations could be large in the nucleon Roper
resonance and the $\Lambda(1405)$ \cite{kimu00,kreh00}. In this case,
these two states could not be described by a simple $q^3$ model as the
one studied here.

\begin{figure}
\includegraphics*[height=8cm]{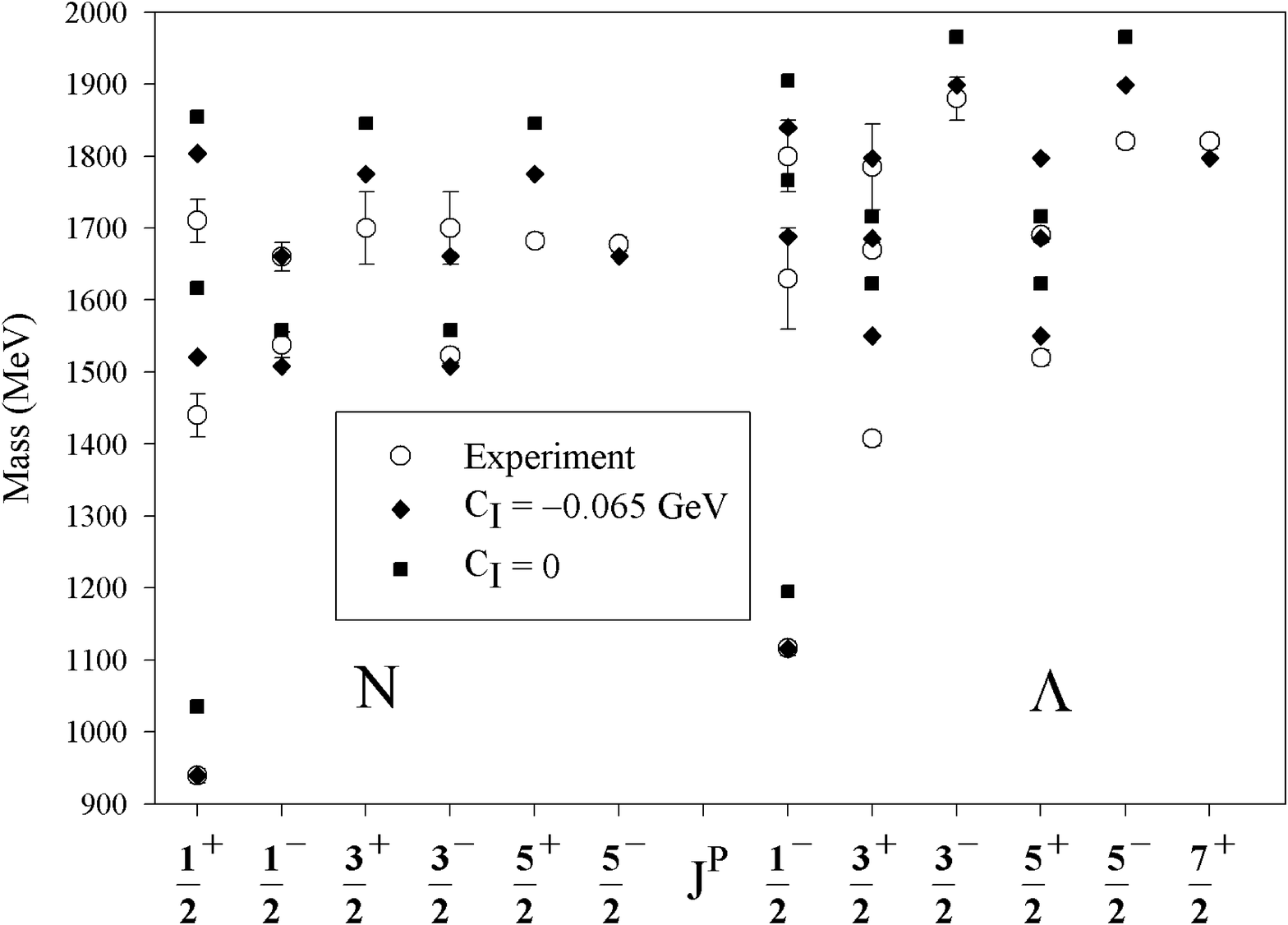}
\caption{Effect of the additional term $C_{\text{I}}$ in the instanton
induced interaction in the baryon sector for the $N$ and $\Lambda$
baryons. The two samples of masses presented are obtained with the same
parameters, except the value of $C_{\text{I}}$.}
\label{fig7}
\end{figure}

The Fig.~\ref{fig7} gives an illustration of the action of
$C_{\text{I}}$ on the energy levels of the $N$ and $\Lambda$ families
(where it plays a major role). For instance, with $C_{\text{I}}=0$, the
$\Delta$ and $\Omega$ baryons are well reproduced, but the masses of
the $N$ and $\Lambda$ sectors are tens of MeV too high. When
$C_{\text{I}}=-65$ MeV, these last baryons are more nicely described
while the states of the $\Delta$ and $\Omega$ sectors remain at the same
place.

\section{Concluding remarks}
\label{sec:rem}

In Ref.~\cite{brau98}, we showed that a simple semirelativistic model
based on the spinless Salpeter equation with a Cornell potential
supplemented by instanton induced forces is able to describe correctly
light meson spectra, provided the quarks are considered as effective
degrees of freedom with a finite size and a constituent mass. The wave
function was partially tested by calculating the electromagnetic mass
splitting.

In Ref.~\cite{sema01}, we showed that this simple semirelativistic model
extended to treat three constituent quarks leads to a rather good
description of the light baryon spectra. However, this model does not
provide a correct unified description of both meson and baryon
spectra. Two different sets of parameters were necessary for such a
calculation; the natural link between meson and baryon was then broken.

In this work, we present a simple extension of these models which allows
a good consistent description of both meson and baryon spectra. An
interaction with one free parameter is added to the instanton induced
interaction in the baryon sector solely. This is, up to now, a purely
phenomenological constant term with the same projector structure than
the leading term of the instanton induced interaction in the baryon
sector.

The spectra calculated with the model defined in Sec.~\ref{sec:model}
are globally better than those obtained with other models built to
describe both mesons and baryons. Obviously, better spectra can be found
in the literature but they are obtained with models dedicated to only
one of these hadron families. These encouraging results incite to
perform further tests of our model by calculating other observables, for
example, electromagnetic mass splittings, electromagnetic form factors
and decay widths. Such a work is in progress \cite{brau01}. The model
could also be extended to heavy mesons and baryons but in this case
another spin-dependent interaction is needed since the instanton induced
interaction does not act in these sectors.




\clearpage

\begingroup
\squeezetable
\begin{table}
\protect\caption{Centers of gravity (c.o.g.)\ of $L$ and $I$ multiplets
for mesons chosen to fix the parameters of the model (the minimal
uncertainty is fixed at 10 MeV, see Ref.~\protect \cite{brau98}). The
values of the c.o.g.\ and their corresponding errors are given by
formula (33) of Ref.~\protect\cite{brau98}. The symbol ``mf ''
means ``mixed flavor''. A meson name used to represent a multiplet in
Figs.~\protect\ref{fig1}, \protect\ref{fig2}, and \protect\ref{fig3} is
underlined.}
\label{tab1}
\begin{ruledtabular}
\begin{tabular}{lccccc}
State & Flavor & $I$ & $J^{P(C)}$ & $N\ ^{2S+1}L_J$ & c.o.g. (GeV) \\
\hline
$\underline{\pi}$           & $n\bar{n}$ & 1   & $0^{-+}$ & $1\
^{1}S_{0}$ &
0.138$\pm$0.010 \\
$\omega$                    & $n\bar{n}$ & 0   & $1^{--}$ & $1\
^{3}S_{1}$ &
0.772$\pm$0.010 \\
$\underline{\rho}$          & $n\bar{n}$ & 1   & $1^{--}$ & $1\
^{3}S_{1}$ &
\\
$h_1(1170)$                 & $n\bar{n}$ & 0   & $1^{+-}$ & $1\
^{1}P_{1}$ &
1.265$\pm$0.013 \\
$b_1(1235)$                 & $n\bar{n}$ & 1   & $1^{+-}$ & $1\
^{1}P_{1}$ &
\\
$f_1(1285)$                 & $n\bar{n}$ & 0   & $1^{++}$ & $1\
^{3}P_{1}$ &
\\
$a_1(1260)$                 & $n\bar{n}$ & 1   & $1^{++}$ & $1\
^{3}P_{1}$ &
\\
$f_2(1270)$                 & $n\bar{n}$ & 0   & $2^{++}$ & $1\
^{3}P_{2}$ &
\\
$\underline{a_2(1320)}$     & $n\bar{n}$ & 1   & $2^{++}$ & $1\
^{3}P_{2}$ &
\\
$\pi_2(1670)$               & $n\bar{n}$ & 1   & $2^{-+}$ & $1\
^{1}D_{2}$ &
1.681$\pm$0.012 \\
$\omega(1600)$              & $n\bar{n}$ & 0   & $1^{--}$ & $1\
^{3}D_{1}$ &
\\
$\rho(1700)$                & $n\bar{n}$ & 1   & $1^{--}$ & $1\
^{3}D_{1}$ &
\\
$\omega_3(1670)$            & $n\bar{n}$ & 0   & $3^{--}$ & $1\
^{3}D_{3}$ &
\\
$\underline{\rho_3(1690)}$  & $n\bar{n}$ & 1   & $3^{--}$ & $1\
^{3}D_{3}$ &
\\
$f_4(2050)$                 & $n\bar{n}$ & 0   & $4^{++}$ & $1\
^{3}F_{4}$ &
2.039$\pm$0.022 \\
$\underline{a_4(2040)}$     & $n\bar{n}$ & 1   & $4^{++}$ & $1\
^{3}F_{4}$ &
\\
$\underline{\pi(1300)}$     & $n\bar{n}$ & 1   & $0^{-+}$ & $2\
^{1}S_{0}$ &
1.300$\pm$0.100 \\
$\omega(1420)$              & $n\bar{n}$ & 0   & $1^{--}$ & $2\
^{3}S_{1}$ &
1.454$\pm$0.026 \\
$\underline{\rho(1450)}$    & $n\bar{n}$ & 1   & $1^{--}$ & $2\
^{3}S_{1}$ &
\\
$\underline{K}$             & $\bar{s}n$ & 1/2 & $0^{-}$ & $1\
^{1}S_{0}$ &
0.496$\pm$0.010 \\
$\underline{K^*(892)}$      & $\bar{s}n$ & 1/2 & $1^{-}$ & $1\
^{3}S_{1}$ &
0.892$\pm$0.010 \\
$K_1(1270)$                 & $\bar{s}n$ & 1/2 & $1^{+}$ & $1\
^{1}P_{1}$ &
1.382$\pm$0.010 \\
$K_0^*(1430)$               & $\bar{s}n$ & 1/2 & $0^{+}$ & $1\
^{3}P_{0}$ &
\\
$K_1(1400)$                 & $\bar{s}n$ & 1/2 & $1^{+}$ & $1\
^{3}P_{1}$ &
\\
$\underline{K_2^*(1430)}$   & $\bar{s}n$ & 1/2 & $2^{+}$ & $1\
^{3}P_{2}$ &
\\
$K_2(1770)$                 & $\bar{s}n$ & 1/2 & $2^{-}$ & $1\
^{1}D_{2}$ &
1.774$\pm$0.012 \\
$K^*(1680)$                 & $\bar{s}n$ & 1/2 & $1^{-}$ & $1\
^{3}D_{1}$ &
\\
$K_2(1820)$                 & $\bar{s}n$ & 1/2 & $2^{-}$ & $1\
^{3}D_{2}$ &
\\
$\underline{K_3^*(1780)}$   & $\bar{s}n$ & 1/2 & $3^{-}$ & $1\
^{3}D_{3}$ &
\\
$\underline{\phi}$          & $s\bar{s}$ & 0   & $1^{--}$ & $1\
^{3}S_{1}$ &
1.019$\pm$0.010 \\
$h_1(1380)$                 & $s\bar{s}$ & 0   & $1^{+-}$ & $1\
^{1}P_{1}$ &
1.482$\pm$0.010 \\
$f_1(1510)$                 & $s\bar{s}$ & 0   & $1^{++}$ & $1\
^{3}P_{1}$ &
\\
$\underline{f_2'(1525)}$    & $s\bar{s}$ & 0   & $2^{++}$ & $1\
^{3}P_{2}$ &
\\
$\underline{\phi_3(1850)}$  & $s\bar{s}$ & 0   & $3^{--}$ & $1\
^{3}D_{3}$ &
1.854$\pm$0.010 \\
$\underline{\phi(1680)}$    & $s\bar{s}$ & 0   & $1^{--}$ & $2\
^{3}S_{1}$ &
1.680$\pm$0.020 \\
$\underline{f_2(2010)}$     & $s\bar{s}$ & 0   & $2^{++}$ & $2\
^{3}P_{2}$ &
2.011$\pm$0.080 \\
$\underline{\eta}$          & mf       & 0   & $0^{-+}$ & $1\ ^{1}S_{0}$
&
0.547$\pm$0.010 \\
$\underline{\eta'}$         & mf       & 0   & $0^{-+}$ & $1\ ^{1}S_{0}$
&
0.958$\pm$0.010
\end{tabular}
\end{ruledtabular}
\end{table}
\endgroup

\begin{table}
\protect\caption{Quantum numbers and masses (the minimal uncertainty is
fixed at 10 MeV, see Ref.~\protect\cite{brau98}) of the baryons chosen
in the fit of parameters.}
\label{tab2}
\begin{ruledtabular}
\begin{tabular}{lccd}
Baryon         & $I$           & $J^{P}$           & \multicolumn{1}
{r}{\text{Mass (GeV)}}
\\
\hline
$N$            & $\frac{1}{2}$ & $\frac{1}{2}^{+}$ & 0.939\pm 0.010 \\
$N(1440)$      & $\frac{1}{2}$ & $\frac{1}{2}^{+}$ & 1.450\pm 0.020 \\
$\Delta$       & $\frac{3}{2}$ & $\frac{3}{2}^{+}$ & 1.232\pm 0.010 \\
$N(1535)$      & $\frac{1}{2}$ & $\frac{1}{2}^{-}$ & 1.537\pm 0.018 \\
$\Lambda$      & 0             & $\frac{1}{2}^{+}$ & 1.116\pm 0.010 \\
$\Sigma$       & 1             & $\frac{1}{2}^{+}$ & 1.193\pm 0.010 \\
$\Sigma^*$     & 1             & $\frac{3}{2}^{+}$ & 1.385\pm 0.010 \\
$\Xi$          & $\frac{1}{2}$ & $\frac{1}{2}^{+}$ & 1.315\pm 0.010 \\
$\Xi^*$        & $\frac{1}{2}$ & $\frac{3}{2}^{+}$ & 1.530\pm 0.010 \\
$\Omega$       & 0             & $\frac{3}{2}^{+}$ & 1.672\pm 0.010 \\
\end{tabular}
\end{ruledtabular}
\end{table}

\begin{table}
\protect\caption{List of parameters of the model. The column
``Meson-Baryon'' contains the parameter values used to compute the meson
and baryon spectra presented in
Fig.~\protect\ref{fig1}-\protect\ref{fig6}. When available, the expected
value of a parameter is also given in the column ``Exp.''. The values of
the quantities $m_n$, $m_s$, $g$, and $g'$ computed with these
parameters are also indicated at the end.}
\label{tab3}
\begin{ruledtabular}
\begin{tabular}{lldc}
Parameters                       & Unit        & \multicolumn{1}
{r}{\text{Meson-Baryon}} &
Exp. \\
\hline
$m^0_n$                          & GeV       & 0.005 &
0.001-0.009 \protect\cite{pdg} \\
$m^0_s$                          & GeV       & 0.167 &
0.075-0.170 \protect\cite{pdg}\\
$\Lambda$                        & GeV       & 0.216 &
0.208$^{+0.025}_{-0.023}$\protect\cite{pdg} \\
$\langle\bar{n}n\rangle$         & GeV$^3$   & -(0.225)^3 &
$(-0.225\pm 0.025)^3$\protect\cite{rein85} \\
$\langle\bar{s}s\rangle/\langle\bar{n}n\rangle$ &  & 0.800 &
$0.8 \pm 0.1$\cite{rein85} \\
$\epsilon$                       &           & 0.000 & 0-1
\protect\cite{brau98} \\
$a$                              & GeV$^2$   & 0.210 &
$0.20 \pm 0.03$ \cite{mich96} \\
$\kappa$                         &            & 0.525 & \\
$C_{\text{M}}$                   & GeV       & -0.691 & \\
$C_{\text{B}}$                   & GeV       & -0.033 & \\
$C_{\text{I}}$                   & GeV       & -0.065 & \\
$\gamma_{n}$                     & GeV$^{-1}$ & 0.779 & \\
$\gamma_{s}$                     & GeV$^{-1}$ & 0.566 & \\
$\delta_{n}$                     & GeV       & 0.186 & \\
$\delta_{s}$                     & GeV       & 0.279 & \\
\hline
$m_n$                            & GeV       & 0.236 & \\
$m_s$                            & GeV       & 0.472 & \\
$g$                              & GeV$^{-2}$   & 2.906 & \\
$g'$                             & GeV$^{-2}$   & 1.710 & \\
\end{tabular}
\end{ruledtabular}
\end{table}

\clearpage

\end{document}